\documentclass[11pt,a4paper]{article}
\usepackage{amsmath,amssymb,bm}
\usepackage{cite}
\usepackage{cite}
\begin{document}

\title{New perturbation theory representation 
of the conformal symmetry braking effects in gauge 
quantum field theory models}

\author{A.~L.~Kataev$^a$\footnote{{\bf e-mail}: kataev@ms2.inr.ac.ru}~~and 
S.~V.~Mikhailov$^b$\footnote{{\bf e-mail}: mikhs@theor.jinr.ru}\\
$^a$ \small {\em Institute for Nuclear 
Research of the Academy of Sciences of Russia}, \\  
\small{\em 117312, Moscow, Russia }\\ 
$^b$ \small {\em Bogoliubov Laboratory of Theoretical Physics, 
JINR}\\
\small{\em 141980 Dubna, Russia}
}
\date{ }
\maketitle

\begin{abstract}
\noindent
We propose a hypothesis on the detailed structure for the representation of 
the conformal symmetry breaking term in
the basic Crewther relation generalized in the perturbation theory 
framework in QCD renormalized in the ${\rm \overline{MS}}$
scheme. We establish the validity of this representation in the 
$O(\alpha_s^4)$ approximation. 
Using the variant of the generalized Crewther relation formulated here allows finding relations between specific 
contributions to the QCD perturbation series coefficients for the flavor nonsinglet part of the Adler function 
$D^{ns}_A$  for the electron-positron annihilation in hadrons and to the perturbation series coefficients for 
the Bjorken sum rule $S_\text{Bjp}$ for the polarized
deep-inelastic lepton-nucleon scattering. We find new relations between the $\alpha_s^4$  coefficients 
of $D^{ns}_A$ and 
$S_\text{Bjp}$.
Satisfaction of one of them serves as an additional theoretical verification of the recent computer analytic
calculations of the terms of order $\alpha_s^4$  in the expressions for these two quantities.
\end{abstract} 


\textbf{1.} 
The conformal symmetry is basic for important theoretical studies 
in various massless quantum field models 
\cite{1}, \cite{2} including QED \cite{3} and QCD (see Sec. 5 in \cite{4}). 
Using this symmetry in studying the axialvector-vector-vector (AVV) triangle
amplitude allowed establishing the fundamental relation between important 
characteristics of different inclusive processes \cite{5}.
The characteristics investigated in \cite{5} were the normalized expression 
for the flavor nonsinglet part $D_A^{ns}$ of the Adler function
$D_A$ for the $e^+e^-$- annihilation process in hadrons and the nonsinglet 
coefficient function $C^{Bjp}$ of the Bjorken sum rule $S_{Bjp}$ for the
process of deep-inelastic scattering (DIS) of polarized leptons on nucleons, 
which also enters the nonsinglet part of the Ellis-
Jaffe sum rule for the DIS of polarized leptons on nucleons.

The basic Crewther relation was soon applied in \cite{6} to the model case where 
diagrams with lepton insertions on the
internal photon lines are not taken into account in QED. 
The relation is also applicable in the imaginary conformally invariant
limit of QCD. 
In these cases, it has the form
\begin{equation}
\label{1}
D \times  C^{Bjp}|_{ci} =1,   
\end{equation}
where the quantities in the left-hand side are defined as
\begin{eqnarray}
\label{2}
D_A^{ns}(a_s)&=& \bigg(N_c \sum_f Q_f^2\bigg)D(a_s)  \\
\label{3}
S_{Bjp}(a_s)&=&\bigg(\frac{1}{6}\frac{g_a}{g_V}\bigg)C^{Bjp}(a_s)
\end{eqnarray}

It is known that the conformal symmetry is broken in the models of 
quantum field theory by renormalization of
charges. These renormalizations lead to the existence of nonzero 
renormalization group (RG) $\beta$-functions (see \cite{7}
for a detailed exposition). Moreover, the factor $\beta(a_s)/a_s$, 
where $a_s = \alpha_s /\pi$, appears as the result of renormalization of
the trace of the energy-momentum tensor. This property was outlined 
theoretically in \cite{8} and demonstrated
explicitly in  \cite{9,10,11,12}; it is connected with the existence of the conformal 
anomaly. Before \cite{13} appeared, the
existence of a generalization of the basic Crewther relation \cite{5} to the case 
of gauge theories with fermions, like
QED and QCD, in higher orders of the perturbation theory (PT) with explicit 
manifestations of renormalizations of
the coupling constants was unclear. 
The $SU(N_c)$ group factors were classified in \cite{13}; these factors arise when 
the QCD PT series for the function $D^{ns}$ in the $O(a_s^3)$ 
approximation (obtained analytically in \cite{14} and later in \cite{15} in
the ${\rm \overline{MS}}$ scheme) is multiplied by the analogous approximation 
for the function $C^{Bjp}$, known at that time from the
calculations in \cite{16}. The studies performed in \cite{13} allowed finding 
an additional contribution to the right-hand side
of (\ref{1}):
\begin{equation}
\label{4}
D(a_s)C^{Bjp}(a_s)= 1 + \Delta_{csb}(a_s)   
\end{equation}
In the third order of the PT, the ``Crewther unity" is modified by the 
conformal symmetry breaking term $\Delta_{csb}$, 
which is expressed as
\begin{equation}
\label{BK}
 \Delta_{\rm{csb}}(a_s)=\bigg(\frac{\beta(a_s)}{a_s}\bigg)
P(a_s)= \bigg(\frac{\beta(a_s)}{a_s}\bigg) \sum_{m\geq 1}  K_{m} a_s^m \,.
\end{equation}

In this order of the PT, the scheme-independent two-loop RG 
$\beta$-function appears in the factor $\beta(a_s)/a_s$, and the
coefficients $K_1$ and $K_2$ determined in \cite{13} appear in the factor $P(a_s)$. 
Moreover, $K_2$ depends on the quadratic
Casimir operators $C_{F}$ and $C_A$ of the $SU(N_c)$ gauge group and on 
the number $n_f$ of fermion flavors.

The discovery of this QCD generalization of the Crewther relation in the 
third-order PT \cite{13} in the ${\rm \overline{MS}}$
scheme with the factor $\beta(a_s)/a_s$ was the first independent theoretical 
indication of the validity of computer analytic
calculations of the $O(a_s^3)$  corrections in the PT series for the 
$D$ functions \cite{14}, \cite{15} and of the analogous
calculations of the third term in the PT series for $C^{Bjp}$ \cite{16}. 
We note that this ``cross-checking" theoretical
indication was later confirmed by independent computer calculations of the 
contributions of the order $a_s^3$ to the $D$
function in \cite{17} using a different theoretical approach.
The next-to-leading PT corrections to the $D$ function were previously evaluated 
analytically in \cite{18} and
numerically in \cite{19}. These results were soon confirmed analytically \cite{20}. 
In the case of $C^{Bjp}$, corrections of the
same PT order were obtained in \cite{21} and later confirmed in \cite{22} using a 
different symbolic computational technique.

To explain the origin of the effect of the factorization $\beta(a_s)/a_s$ in 
Eq. (\ref{BK}), the operator product expansion
method in the momentum space was applied in \cite{23} to the triangle diagram of 
the AVV currents (see \cite{24} for a
more detailed discussion). Also in \cite{23}, 
arguments were presented for the absence of an inconsistency between the
one-loop nature of the axial anomaly \cite{25} and the multiloop structure of 
the QCD generalization of the Crewther
relation, based on their relation to different form factors in the AVV 
triangle diagram \cite{23}. The possibility that the
multiloop factor $\beta(a_s)/a_s$ in the conformal symmetry breaking term 
in Eq. (\ref{BK}) is factorable in all PT orders \cite{23} with
the coefficients $K_m$ of the polynomial $P(a_s)$ unfixed in the presented 
considerations was also indicated. The
considerations imply the application of the ${\rm \overline{ MS}}$ 
scheme in which the coefficient functions of the leading operators
in the operator product expansion method can be explicitly defined. 
The second conclusion in \cite{23} was proved in the coordinate space in \cite{26} and 
was previously discussed in \cite{27} but was published only recently \cite{4}. 
This variant of the QCD generalization of the Crewther
relation in the ${\rm \overline{MS}}$ scheme \cite{13} was considered from a more 
phenomenological standpoint in \cite{28} and \cite{29}, where
the characteristic energy scales were fixed by applying the multiloop 
version (developed in \cite{30}) of the Brodsky-
Lepage-Mackenzie approach \cite{31} supplemented by the procedure for constructing 
the ``commensurate scale relations" in \cite{32}. As a result, the 
``Crewther unity" was restored in the right-hand side of (\ref{4}) at the $O(a_s^3)$ 
level by absorbing the conformal symmetry breaking term into the energy scale 
of the effective charge for the $D$ function,
which is equivalent to choosing a certain scale in the invariant charge for 
the $D$ function and including the
conformal symmetry breaking term in the energy scale of the effective 
charge for $C^{Bjp}$ \cite{29}.

In the case of the $SU(N_c)$ group, the $O(a_s^4)$ 
corrections to the functions $D^{ns}$ and $C^{Bjp}(a_s)$ are known thanks to
the recent analytic calculation in the $\overline{MS}$ scheme \cite{33}. 
These calculations with the three-loop analytic contributions
to the QCD $\beta$-function in this scheme taken into account \cite{34}, \cite{35} 
allowed fixing the coefficient $K_3$ of the
polynomial $P(a_s)$ in expression (\ref{BK}) and demonstrating the validity of 
the results in \cite{28} with the $O(a_s^4)$ PT
contributions taken into account. We note that the explicit results for 
$C^{Bjp}$ confirmed the expressions for the $\zeta_3$-
containing QED contributions to the Bjorken sum rule that appeared first 
in  fourth-order QED PT corrections to
the function $D^{ns}$ \cite{36}. This term in the Bjorken sum rule was previously 
obtained in \cite{37} from the results in \cite{36},
arguments based on the conformal symmetry and the basic Crewther relation
\footnote{The arguments for the 
possibility of explaining the appearance at this PT level of 
$\zeta_3$ term, untypical for previously known diagram 
contributions characterizing the photon function of the QED vacuum polarization 
without fermion loop insertions into internal photon lines, were 
presented in \cite{38}}.  
The found agreement was the first
confirmation of the validity of the calculations in \cite{33}.

Our main purpose here is to justify the detailed representation of the 
generalized Crewther relation at the $a_s^4$
level previously proposed in \cite{39}. Its new feature is writing the 
right-hand side of Eq. (\ref{BK}) in the form of a double
power expansion in which the first expansion ``parameter" is the function 
$\beta(a_s)/a_s$ and the second is the coupling
constant $a_s$, namely,
\begin{eqnarray}
\Delta_{\rm{csb}}(a_s) = \sum_{n\geq 1}\bigg(\frac{\beta(a_s)}{a_s}\bigg)^n
{\cal P}_n(a_s) &=& \sum_{n\geq 1}\sum_{r\geq 1}\bigg(\frac{\beta(a_s)}{a_s}\bigg)^n P_n^{(r)}a_s^r = \nonumber\\
&=& \sum_{n\geq 1}\sum_{r\geq 1}\bigg(\frac{\beta(a_s)}{a_s}\bigg)^n P_n^{(r)} [{\rm{k,m}}] C_F^{\rm{k}} C_A^{\rm{m}} 
a_s^r \label{KM1}, %, \label{KM2}
\end{eqnarray}
where $\rm{k + m}$ = $r$ and the coefficients $P_n^{(r)}[{\rm{k,m}}]$
 contain rational fractions and Riemann $\zeta$-functions of odd
arguments. In contrast to the coefficients of the polynomial 
$P(a_s)$ in Eq. (\ref{BK}), the coefficients of ${\cal P}_n(a_s)$ in  Eq. (\ref{KM1}) are
independent of the number ${\rm n_f}$ of quark flavors.

\textbf{2.} 
We consider the PT series for the nonsinglet part of the Adler function $D$ 
and the coefficient function $C^{Bjp}$
for the Bjorken sum rule respectively defined in (\ref{2}) and (\ref{3}) and 
normalized 
to unity:
\begin{equation}
\label{DAp}
D=1 + \sum_{n=1} d_n~a_s^n,~~~C^{Bjp}=1 + \sum_{l=1} c_l~a_s^l~. 
\end{equation} 

Explicit expressions for $d_1$, $d_2$, $d_3$ and $c_1$, $c_2$, $c_3$ 
in terms of the $SU(N_c)$ group factors are well known (see, e.g.,
\cite{14}, \cite{16}). In the concrete case of the $SU(3)$ group, the fourth coefficient 
$d_4$ of the $D$ function was evaluated
analytically in \cite{40}. This result was recently generalized to the case of an 
arbitrary color group $SU(N_c)$ in \cite{33}. The
analogous coefficient $c_4$ for $C^{Bjp}$, 
also calculated in \cite{33} (see the supplemental file to
the electronic preprint version of \cite{33}), is \footnote{
 We recovered this expression, which agrees with the result contained 
in the electronic supplement to the preprint of \cite{33},
from its text in which the result for $1/C^{Bjp}$ was presented}
\begin{eqnarray}
 c_4&=&
\left[
-\frac{3}{16} 
+\frac{1}{4}\zeta_3
+\frac{5}{4}\zeta_5
\right]{\rm \frac{d_F^{abcd}d_A^{abcd}}{d_R}} +
\left[
\frac{13}{16} 
+ \zeta_{3}
-\frac{5}{2}\zeta_5
\right]{\rm \frac{d_F^{abcd}d_F^{abcd}}{d_R}\rm n_f}
-
\left[\frac{4823}{2048}+\frac{3}{8}\zeta_3\right]{\rm C_F^4}
 \nonumber \\
&+&\left[\frac{839}{2304}+\frac{451}{96}\zeta_3
-\frac{145}{24}\zeta_5\right]{\rm C_F^3T_Fn_f}
+\left[-\frac{265}{576}
+\frac{29}{24}\zeta_3\right]{\rm C_F^2T_F^2n_f^2}
+\left[\frac{605}{972}\right]{\rm C_FT_F^3n_f^3}\nonumber  \\ 
&+&\left
[-\frac{3707}{4608}-\frac{971}{96}\zeta_3+\frac{1045}{48}\zeta_5\right]{\rm C_F^3C_A}
+\left[-\frac{87403}{13824}-\frac{1289}{144}\zeta_3
+\frac{275}{144}\zeta_5+\frac{35}{4}\zeta_7\right]{\rm C_F^2C_AT_Fn_f} \nonumber  \\
&+&\left[-\frac{165283}{20736}-
\frac{43}{144}\zeta_3+\frac{5}{12}\zeta_5
-\frac{1}{6}\zeta_3^2\right]{\rm C_FC_A T_F^2 n_f^2} \nonumber \\ 
&+&\left[\frac{1071641}{55296}
+\frac{1591}{144}\zeta_3-\frac{1375}{144}\zeta_5
-\frac{385}{16}\zeta_7\right]{\rm C_F^2C_A^2} \nonumber \\ 
&+&\left[\frac{1238827}{41472}
+\frac{59}{64}\zeta_3-\frac{1855}{288}\zeta_5
+\frac{11}{12}\zeta_3^2-\frac{35}{16}\zeta_7\right]{\rm C_FC_A^2T_Fn_f}  \nonumber \\ 
&+&\left[-\frac{8004277}{248832}
+\frac{1069}{576}\zeta_3+\frac{12545}{1152}\zeta_5-
\frac{121}{96}\zeta_3^2+\frac{385}{64}\zeta_7\right]{\rm C_FC_A^3}~.
\label{eq:c_4}
\end{eqnarray}
where $\zeta_{2q+1}=\sum_{k=1}^{\infty}(1/k)^{2q+1}$ is the 
Riemann function of an  odd  argument. In the fundamental 
representation of $SU(N_c)$, the group factors are defined as 
 ${\rm C_F}=(N_c^2-1)/(2N_c)$, 
${\rm C_A}=N_c$, ${\rm T_F}=1/2$,  
${\rm d_F^{abcd}d_A^{abcd}/d_R=N_c(N_c^2+6)/18}$,and   
${\rm  d_F^{abcd}d_F^{abcd}/{d_R}=(N_c^4-6N_c^2+18)/(36N_c^2)}$.
For the $SU(3)$ group, which corresponds to the QCD 
case, we have  ${\rm C_F}=4/3$, ${\rm C_A}=3$, ${\rm d_R}=3$
and  ${\rm  d_F^{abcd}d_A^{abcd}} = 15/2$, 
${\rm d_F^{abcd}d_F^{abcd}} = {5}/{12}$. 

A strong verification of the self-consistency
of the results obtained in \cite{40} and \cite{33} follows from the validity of 
QCD-generalized Crewther relation (\ref{BK}) after
the $O(a_s^4)$ contributions to the left-hand side of (\ref{4}) evaluated in the 
${\rm \overline{MS}}$-scheme are taken into account. We recall
that the existence of this generalization with the factored multiplier 
$\beta(a_s)/a_s$ is not accidental. It was discovered in
the preceding PT order \cite{13} and proved in all orders in \cite{26}. It was 
shown in \cite{13} that the coefficients of the
polynomial $P(a_s)$ in (\ref{BK}) in the third order of the PT can be 
expressed as
\begin{eqnarray}
\label{10}
 K_{1}&=& K_{1}[1,0,0]{\rm C_F}, \nonumber \\ 
\label{11} K_{2}&=&K_{2}[2,0,0]{\rm C_F^2}+K_{2}[1,1,0]{\rm C_F C_A} + K_{2}[1,0,1]
 {\rm  C_F T_F n_f},
 \end{eqnarray}
The fourth-order PT calculations in \cite{33} lead to the fixation of the 
third term in the polynomial $P(a_s)$ in the form of
a sum of six terms proportional to the Casimir operators of the 
$SU(N_c)$ group times the number ${\rm n_f}$ of fermion flavors:
\begin{eqnarray}
\label{eq:K_3}
K_3&=& K_3[3,0,0]{\rm C_F^3}+K_3[2,1,0]{\rm C_F^2 C_A}+K_3[1,2,0]{\rm C_F C_A^2}  
+K_3[2,0,1]{\rm C_F^2 T_F n_f} \nonumber \\
&&+K_3[1,1,1]{\rm C_F C_AT_Fn_f}  
+K_3[1,0,2]{\rm C_F (T_F n_f)^2}. 
\end{eqnarray} 
The analytic expression for the last coefficient 
$K_3[1,0,2]$ in (10) coincides with the result in \cite{13} obtained when
calculating analogous coefficients generated in higher PT orders by 
multiplying the contributions to the functions
$D(a_s)$ and $C^{Bjp}$ of the diagrams with a large number of 
one-loop fermion insertions into the internal gluon lines.

In correspondence with the structure of the term $\Delta_{csb}(a_s)$ 
in (\ref{BK}) and (\ref{KM1}), we need concrete values of the coefficients
of the RG $\beta$-function in the ${\rm \overline{MS}}$- scheme
\begin{equation}
\label{bf}
\mu^2 \frac{d}{d \mu^2} a_s=
\beta(a_s)=- a_s^2\left(\beta_0 + \beta_1 a_s + \beta_2 a_s^2 \right),
\end{equation}
found in the three-loop approximation in \cite{34} and confirmed in \cite{35}.  
The coefficients $\beta_i$ can be expressed in the
forms
\begin{eqnarray}
\beta_0&=&\beta_0[0,1,0]{\rm C_A}+\beta_0 [0,0,1]{\rm T_F n_f},\nonumber \\ %\label{beta0}
\beta_1&=& \beta_1[0,2,0]{\rm C_A^2}+\beta_1[0,1,1]{\rm C_A T_F n_f}
+\beta_1[1,0,1]{\rm  C_F T_F n_f} \label{beta1} \nonumber \\ 
\beta_2&=& \beta_2[0,3,0]{\rm C_A^3}+\beta_2[0,2,1]{\rm C_A^2T_Fn_f}+
\beta_2[1,1,1]{\rm C_FC_AT_Fn_f} \nonumber \\
&&+\beta_2[0,1,2]{\rm C_AT_F^2n_f^2}+\beta_2[2,0,1]{\rm C_F^2T_Fn_f}+\beta_2[1,0,2]{\rm C_FT_F^2n_f^2}~\label{beta2}, 
\end{eqnarray}
with the elements $\beta_j [... ]$:
\begin{eqnarray}
%\label{17}
\beta_0[0,1,0]&=&\frac{11}{12}~~,~\beta_0[0,0,1]=-\frac{1}{3}, \nonumber\\
\beta_1 [0,2,0]&=& \frac{17}{24}~~, ~\beta_1 [0,1,1]=-\frac{5}{12}~,~
\beta_1[1,0,1]=-\frac{1}{4}, \nonumber \\%\label{18}
\beta_2[0,3,0]&=&\frac{2857}{3456}~,~\beta_2[0,2,1]=-\frac{1415}{1728}~,~
\beta_2[1,1,1]=-\frac{205}{576}  \nonumber \\%\label{19}
\beta_2[0,1,2]&=&\frac{79}{864}~,~\beta_2[2,0,1]=\frac{1}{32}~,~
\beta_2[1,0,2]=\frac{11}{144}~. \label{20}
\end{eqnarray}

\textbf{3.} 
We now consider the issue of the uniqueness of a detailed generalization 
of Crewther relation (\ref{KM1}) in powers
of the $\beta$-function. We here present additional arguments for our 
assumption that such a generalization exists (see
\cite{39}) and justify it using the results of the fourth-order PT approximation 
for (\ref{4}) and (\ref{BK}) obtained in \cite{33}.
The derivation of the detailed generalization of the Crewther relation in 
the ${\rm \overline{MS}}$ scheme is based on the
requirement that the coefficients of the polynomials 
 ${\cal P}_n$ in (\ref{KM1}) should be independent of the $\beta$-function 
coefficients
and consequently independent of the number ${\rm n_f}$ of fermion flavors. 
This property can be realized by passing from
representation (\ref{BK}) with the single factored $\beta$-function in the 
expression for the conformal symmetry breaking term
$\Delta_{csb}(a_s)$ in (\ref{4}) to representation (\ref{KM1}) in the form of an expansion 
in powers of $\beta(a_s)/a_s)$. The validity of this form of
writing $\Delta_{csb}(a_s)$ in the fourth PT order was assumed in \cite{39} before 
the publication of the analytic results of
calculations of the $D$-function and $C^{Bjp}(a_s)$ in the $a_s^4$ order 
\cite{33}. To derive it explicitly, we should equate the right-
hand sides of the two representations for $\Delta_{csb}(a_s)$ from (\ref{BK}) 
and (\ref{KM1}) at each order of the expansion in the coupling
constant $a_s$. In the PT approximations we are interested in, the 
coefficients in the right-hand side of (\ref{BK}) are related 
to the analogous contributions to (\ref{KM1})
by the system of linear equations
\begin{eqnarray}
K_1[1,0,0]&=&P_1^{(1)}[1,0], \nonumber \\ 
K_2[2,0,0]&=&P_1^{(2)}[2,0], \nonumber \\  
K_2[1,1,0]&=&P_1^{(2)}[1,1]-\beta_0[0,1,0]P_2^{(1)}[1,0], \nonumber \\ 
K_2[1,0,1]&=&-\beta_0[0,0,1]P_2^{(1)}[1,0], \nonumber ~~\\
K_3[3,0,0]&=& P_1^{(3)}[3,0],  \nonumber \\
K_3[2,1,0]&=&P_1^{(3)}[2,1]-\beta_0[0,1,0] P_2^{(2)}[2,0], \nonumber \\
K_3[1,2,0]&=& P_1^{(3)}[1,2]-\beta_0[0,1,0] P_2^{(2)}[1,1]-\beta_1[0,2,0] 
P_1^{(1)}[1,0]+(\beta_0[0,1,0])^2 P_3^{(1)}[1,0],   \nonumber \\
K_3[2,0,1]&=&-\beta_1[1,0,1]P_2^{(1)}[1,0]-\beta_0[0,0,1]P_2^{(2)}[2,0], 
\nonumber \\
K_3[1,1,1]&=&-\beta_1[0,1,1]P_2^{(1)}[1,0]-\beta_0[0,0,1]P_2^{(2)}[1,1]
+2\beta_0[0,1,0]\beta_0[0,0,1]P_3^{(1)}[1,0], \nonumber  \\
K_3[1,0,2]&=&(\beta_0[0,0,1])^2 P_3^{(1)}[1,0]. \label{syst}
\end{eqnarray}
The unique  solution of  this  system  determines the explicit expressions
for the  three polynomials ${\cal P}_n(a_s)$ 
 with coefficients $P_n^{(r)} [{\rm{k,m}}]$ independent of the number 
of flavours: 
\begin{eqnarray}
{\cal P}_1(a_s)&=&\bigg(-\frac{21}{8}+3\zeta_3 \bigg){\rm C_F} a_s+  
 \bigg [\bigg(\frac{397}{96} + \frac{17}{2}\zeta_3 -15\zeta_5\bigg){\rm C_F^2}+
  \bigg(-\frac{47}{48}+\zeta_3 \bigg){\rm C_FC_A}\bigg ] 
 a_s^2~\nonumber \\  
&  &+\bigg [ \bigg
( \frac{2471}{768}+\frac{61}{8}\zeta_3-\frac{715}{8}\zeta_5
+\frac{315}{4}\zeta_7\bigg){\rm C_F^3}~~ 
  \nonumber \\
&&+\bigg(\frac{16649}{1536}-\frac{11183}{192}\zeta_3+\frac{1015}{24}
\zeta_5-\frac{105}{8}\zeta_7+\frac{99}{4}\zeta_3^2\bigg){\rm C_F^2 C_A} ~~  \nonumber  \\
&& +  \bigg(\frac{2107}{192}+\frac{2503}{72}\zeta_3-\frac{355}{18}
\zeta_5-33\zeta_3^2\bigg){\rm C_FC_A^2}\bigg]a_s^3+O(a_s^4);  \label{P1} 
 \end{eqnarray}
  \begin{eqnarray}
\label{P2}
{\cal P}_2(a_s) &=&   \bigg(\frac{163}{8}
- 19 \zeta_3\bigg){\rm C_F} a_s  
+ \bigg[ \bigg(-\frac{13597}{384}
- \frac{2523}{16}\zeta_3+\frac{375}{2}\zeta_5+27\zeta_3^2\bigg){\rm  C_F^2} 
\nonumber \\ \nonumber 
&&+  \bigg(\frac{1433}{32}-\frac{1}{4}\zeta_3-\frac{85}{2}\zeta_5-
6\zeta_3^2\bigg){\rm C_FC_A}
\bigg]a_s^2  +O(a_s^3);   \\  \nonumber 
 {\cal P}_3(a_s)&=& 
 \bigg(-\frac{307}{2} +\frac{203}{2}\zeta_3+45\zeta_5\bigg){\rm C_F} a_s +O(a_s^2). 
 \end{eqnarray} 

We note that the four-loop term $\beta_3$ 
of the RG $\beta$-function, evaluated analytically in the case of 
$SU(N_c)$ in \cite{41} and
confirmed in \cite{42}, contains three new group structures 
${\rm d_A^{abcd}d_A^{abcd}}$, ${\rm d_F^{abcd}d_A^{abcd}n_f}$
and ${\rm d_F^{abcd}d_F^{abcd}n_f^2}$.
In view of the
factorization of the $\beta$-function in (\ref{BK}) in all PT orders 
(see the proofs in \cite{26}, \cite{4}), we conclude
that the appearance of these extra group terms does not spoil the 
$\beta$-function factorability in (\ref{BK}) and also in the first
term of the sum in (\ref{KM1}).

One more conclusion follows from higher contributions in powers of 
${\rm n_f}$ calculated in \cite{13}, equivalent to
calculating the corrections proportional to higher powers of the first 
coefficient $\beta_0$ of the RG $\beta$-function. These
corrections determine the leading contributions to the 
polynomials ${\cal P}_n(a_s)$ of the new representation 
for $\Delta_{csb}$ in (\ref{KM1}), which have the form
\begin{equation}
\label{Deltan}
{\cal P}_n(a_s)=\frac{S_n}{4^n}3^{(n-1)}{\rm C_F}a_s+O(a_s^2).
\end{equation}
The first nine coefficients $S_n$ , $1\leq n\leq 9$,  
were calculated analytically in \cite{13}.

\textbf{ 4.}~ Representation (\ref{KM1}) can be obtained differently by using 
the $\beta$-expansion formalism of the coefficients of the
PT series (in the ${\rm\overline{MS}}$ scheme) developed in \cite{43}. 
In this approach, instead of the commonly used expansions of the
coefficients in powers of the flavor-dependent factor $T_{F}{\rm n_f}$, 
the quadratic Casimir operators ${\rm C_F}$ and ${\rm C_{A}}$, and the
structure constants of the color group $SU(N_c)$, 
it is proposed to consider expansions of the coefficients 
$d_{n}$ and $c_{n}$ in
powers of the coefficients $\beta_0$, $\beta_1, \dots $ of the 
$\beta$-function with the weight elements $d_n[n_0, n_1,...]$ and 
$c_n[n_0, n_1,...]$.
Their first arguments ($n_0$) determine the powers of the coefficients $\beta_0$ of the elements 
$d_n[\ldots]$ and $c_n[\ldots]$, the second
arguments ($n_1$) give the powers of the coefficients $\beta_1$, and so on. 
The elements $d_n[0, 0,..., 0]$ and $c_n[0, 0,..., 0]$ 
are the contributions ``cleaned" of the charge renormalizations and the 
factors $\beta_i$, whose powers are here equal to zero
($n_i = 0$). These elements coincide with the values of the coefficients 
$d_n$ and $c_n$ in the hypothetical limit with the $\beta$-
function identically equal to zero in all PT orders in QCD. 
This limit corresponds to restoring the conformal
symmetry in the effective quantum field model. 
We regard the transition to this model as a technical trick here. If
all arguments $n_i$ of the elements $d_n[..., m, 0,..., 0]$ and 
$c_n[..., m, 0,..., 0]$ after some index $m$ are zero, 
then we simplify the notation as 
$d_n[..., m, 0,..., 0] = d_n[..., m]$ and $c_n[..., m, 0,..., 0] = 
c_n[..., m]$. The corresponding $\beta$-
representations for the first few coefficients of (\ref{DAp}) are
\begin{eqnarray}
\label{eq:d_2}
d_2&=&\beta_0\,d_2[1]
  + d_2[0]\, ,\\
\label{eq:d_3}  d_3
&=&
  \beta_0^2\,d_3[2]
  + \beta_1\,d_3[0,1]
  + \beta_0\,d_3[1]
  + d_3[0]\, , \\ 
  d_4
   &=& \beta_0^3\, d_4[3]
     + \beta_1\,\beta_0\,d_4[1,1]
     + \beta_2\, d_4[0,0,1]
     + \beta_0^2\,d_4[2]
     + \beta_1\,d_4[0,1]
     + \beta_0\,d_4[1]
     + d_4[0]\,. \label{eq:d_4}
\end{eqnarray}
Analogous representations also hold for higher coefficients 
$d_n$ in the PT series for the $D$ function and for the
coefficients $c_l$ of the PT series for $C^{Bjp}$ given by (\ref{DAp}), and so on. 
We stress that the representations like (\ref{eq:d_2})-(\ref{eq:d_4}) are
unique. The coefficients $d_n[n- 1]$ and $c_n[n-1]$ are identical to the 
terms generated by the chains of one-particle-
reducible one-loop fermion insertions into the gluon propagators and can be 
found, for example, in \cite{13}.
Determining the explicit forms of the other elements is a separate and not 
simple task. Their diagram
representation was discussed in \cite{43}. Below, we consider a way to 
obtain concrete analytic expressions for the
elements of the coefficients $d_n$ and $c_l$ 
up to corrections of the order $a_s^3$.
Expansion (\ref{KM1}) together with (\ref{eq:d_2})-(\ref{eq:d_4}) allows finding the relation between the 
unknown elements of the
fourth-order PT coefficients $d_4$ and $c_4$ and the elements in the 
expressions for the third order of the PT series
(which are presented explicitly below).

By virtue of relation (1) following from the unbroken conformal symmetry 
restored in the hypothetical case at
$\beta_i = 0$, we find an explicit relation between the contributions 
``cleaned" from the charge renormalizations:
\begin{equation}
c_n[0] + d_n[0] + \sum_{l=1}^{n-1} d_l[0] c_{n-l}[0]=0.
\label{eq:CI-PT0}
\end{equation}
The special feature of this recurrence relation is the possibility to express the sum of the 
$n$th-order PT elements in terms of the
analogous elements in the coefficients of lower PT approximations. 
The relation for the ``cleaned" elements $c_4[0]$ and $d_4[0]$ of
the coefficients of the fourth-order PT hence follows:
\begin{equation} 
\label{eq:k4-d4}
c_4[0] + d_4[0]= 2d_1 d_3[0]-3d_1^2d_2[0]+(d_2[0])^2+d_1^4.
\end{equation} 

We note that this equation contains contributions proportional not only to 
the Casimir operators 
${\rm C_F}$ but also to   ${\rm C_A}$. We recall that the projection 
of relation (\ref{eq:k4-d4}) onto the maximum power of ${\rm C_F}$, ${\rm C_F}^4$, 
is equivalent to the relation previously used in \cite{37} to
formulate the proposed verification of the QED result for an analogue of 
$d_4$ first published in \cite{36}.
The explicit expression for $d_3$ in the $\beta$-expansion was obtained 
in \cite{43} thanks to using the analytic result evaluated in \cite{17}
for the contribution to the third coefficient of the PT series for the 
Adler function $D(a_s,{\rm n_f},n_{\tilde{g}})$ with $n_{\tilde{g}}$ gluino 
multiplets when
the contributions from scalar quarks (squarks) are neglected in the 
supersymmetric variant of QCD. At the level of $a_s^2$
corrections, the analytic result for the gluino contributions in \cite{17} 
coincides with the numerical result in \cite{44}, and the gluino
correction of the order $a_s^3$ evaluated analytically in the 
${\rm \overline{ MS}}$ scheme in \cite{17} was confirmed in \cite{45}.
It is easy to obtain the element $d_3[2]$ in (\ref{eq:d_3}). 
Its value can also be extracted from the results in \cite{13}. 
We should then separate the contributions from the terms $\beta_1 d_3[0,1]$ 
and $\beta_0 d_3[1]$ in the expression for $d_3$. 
They are both linear in the number
${\rm n_f}$ of quark flavors. They separate if we use additional degrees of 
freedom, the abovementioned gluino contributions labeled by
the number $n_{\tilde{g}}$ of gluino multiplets.\footnote{ 
We note that the possible existence of a gluino with a mass in the region 
$m_{\tilde{g}}\geq   195~{\rm  GeV}$, lighter than the squark in the minimal
supersymmetric standard model (MSSM), is not excluded by the existing Tevatron data \cite{46} but was recently 
excluded by LHC data. Nevertheless, the joint detailed analysis of the 
available LHC data by the CMS and ATLAS collaborations still does not 
exclude the possible existence of a gluino with a mass in the 
region $m_{\tilde{g}} \geq  400~ {\rm GeV}$ heavier than squarks \cite{47}.}

We can then find the explicit forms of the functions 
${\rm n_f} = {\rm n_f}(\beta_0,\beta_1)$ and $n_{\tilde{g}} = n_{\tilde{g}}(\beta_0,\beta_1)$. 
These expressions can be obtained after taking the gluino contributions to 
the first two coefficients of the $\beta$-functions for this type of 
extension of QCD into account. These two-loop results are known from the 
calculations in \cite{48}. The coefficients of the $\beta$-expansions of 
the terms $d_2$ and $d_3$  defined in (\ref{eq:d_2}) and (\ref{eq:d_3}) were obtained just this 
way in \cite{43}. We here present the results in \cite{43} only slightly changing
the normalization coefficients:
\begin{eqnarray}
d_1=\frac{3}{4}{\rm C_F},~d_2[1]= \bigg(\frac{33}{8} - 3\zeta_3\bigg){\rm C_F},~
d_2[0]= -\frac{3}{32}{\rm C_F^2}  + \frac{1}{16}{\rm C_FC_A},&&
 \\
d_3[2]=\bigg(\frac{151}{6}-19\zeta_3\bigg){\rm C_F},~ d_3[1]=
    \bigg(-\frac{27}{8} - \frac{39}{4}
    \zeta_3 + 15\zeta_5\bigg){\rm C_F^2}
-\bigg(\frac{9}{64} - 5\zeta_3 +\frac{5}{2}\zeta_5\bigg)
{\rm C_FC_A} \label{D-31}, &&\\
d_3[0,1]= \bigg(\frac{101}{16}-6\zeta_3\bigg){\rm C_F}, 
~d_3[0]= - \frac{69}{128}{\rm  C_F^3}+ \frac{71}{64}{\rm C_F^2C_A}+
    \bigg(\frac{523}{768}- \frac{27}{8}\zeta_3\bigg){\rm C_FC_A^2}~. && \label{D-30}
\end{eqnarray} 

We now express the elements $c_3[\ldots ]$ in an analogous form. 
Using (20) to determine $c_3[0]$ and taking the analytic
expression for $d_3[0]$ in (24) into account, we obtain
\begin{eqnarray}
c_1&=&-\frac{3}{4}{\rm  C_F},~ 
c_2[1]= -\frac{3}{2}{\rm C_F},~
c_2[0]= \frac{21}{32}{\rm C_F^2}-\frac{1}{16}{\rm C_FC_A}, \nonumber  \\
c_3[2]&=&-\frac{115}{24}{\rm C_F},~
c_3[1]= \bigg(\frac{83}{24}- \zeta_3\bigg){\rm C_F^2} + 
\bigg(\frac{215}{192}- 6 \zeta_3+\frac{5}{2}\zeta_5\bigg){\rm C_FC_A},
\label{c-31} \\ \nonumber 
c_3[0,1]&=&\bigg(-\frac{59}{16}+3\zeta_3\bigg){\rm C_F},~
c_3[0] = - \frac{3}{128} {\rm C_F^3} -\frac{65}{64} {\rm C_F^2C_A}-
\bigg(   \frac{523}{768} - \frac{27}{8}\zeta_3\bigg){\rm C_FC_A^2}. \label{c-30}
\end{eqnarray}
Expansions similar to (\ref{eq:d_2}),~(\ref{eq:d_3}) were previously 
used in \cite{49} both for the 
Adler function and for the Bjorken sum rule. But
only the terms proportional to powers of $\beta_0$ 
(including its zeroth power) were then taken into account. In general, it is 
more consistent to use the approach in \cite{43}, which prescribes also taking 
the contribution of the two-loop coefficients $\beta_1$ of the RG 
$\beta$-function into account.
Now substituting the corresponding forms (\ref{eq:d_2})-(\ref{eq:d_4}) for 
$d_i$ and $c_i$ in our proposed representation (\ref{KM1}), we obtain the
expressions
\begin{eqnarray} 
\nonumber
{\cal P}_1(a_s)&=& a_s \bigg\{P^{(1)}_1+a_s P^{(2)}_1 + 
a_s^2 P^{(3)}_1 \bigg\} \\  \label{eq:P1} 
&=&- a_s \bigg\{c_2[1] + d_2[1]+a_s\Big(c_3[1] + d_3[1]+ d_1\big(c_{2}[1]-d_{2}[1]\big) \Big) \nonumber  \\ %\nonumber
&& + a_s^2 
\Big(c_4[1]+d_4[1]+d_1
\big(c_3[1]-d_3[1]\big)+d_2[0]c_2[1]+d_2[1]c_2[0]\Big)\bigg\} \\ 
\nonumber 
{\cal P}_2(a_s)&=&a_s \big\{P^{(1)}_2+a_sP_2^{(2)} \big\} \\
\label{eq:P2}
&=&a_s\bigg\{c_3[2] + d_3[2]+a_s\Big( 
c_4[2]+d_4[2]
-d_1(c_3[2]-d_3[2])\Big)\bigg\}  \\
\label{eq:P3}
{\cal P}_3(a_s)&=&a_s  P^{(1)}_3 =
- a_s\big\{c_4[3] + d_4[3]\big\}= a_s{\rm C_F} \bigg(\frac{307}{2}-\frac{203}{2}\zeta_3-45\zeta_5\bigg)
\\  
\label{eq:Pn}
{\cal P}_n(a_s)
& &a_s  P^{(1)}_n =
(-1)^{n-1} a_s\big\{c_n[n-1] + d_n[n-1]\big\}  
\end{eqnarray}
The concrete expression for (\ref{eq:Pn}) is defined in (\ref{Deltan}). 
We stress that the analytic form of formulae (\ref{P1}) obtained previously
acquires a concrete relation to the $\beta$-expansion method (see 
(\ref{eq:P1}-\ref{eq:P3}). 
The elements $d_n[n - 1](c_n[n - 1])$ are defined by the
diagrams containing a single gluon propagator with a chain of 
one-particle-reducible one-loop fermion insertions (so-called
leading renormalon contributions) and can be determined from the results obtained in \cite{13}. 
The elements $d_n[l],~l < n - 1$, are
defined by the diagrams with at least two gluon propagators with both 
containing the one-particle-reducible one-loop fermion
insertions (so-called subleading renormalon contributions). 
The similar classes of diagrams have not yet been evaluated
explicitly. Using main theoretical result (\ref{KM1}), which we have explicitly 
verified in the fourth PT order, we can obtain the relations
between the elements of the $\beta$-expansion coefficients $d_4~(d_n)$ and 
$c_4~(c_n)$. Thus, the first term of the polynomial ${\cal P}_1(a_s)$
 in (\ref{KM1}) is determined by the chain of equations
\begin{eqnarray}
  \label{eq:k_n1-d_n1}
P^{(1)}_1&=& - c_2[1] - d_2[1] =- c_3[0,1] - d_3[0,1]= -c_4[0,0,1]-d_4[0,0,1] = \ldots \nonumber \\
&=&  -c_n[\underbrace{0,0,\ldots, 1}_{n-1}] - d_n[\underbrace{0,0,\ldots, 1}_{n-1}] =
{\rm C_F} \left(-\frac{21}8+3\zeta_3 \right)
 \end{eqnarray}
  The second term $P_1^{(2)}$   of the same polynomial, analytically fixed in 
(\ref{P1}), also relates different elements of
the $\beta$-expansion approach:
\begin{eqnarray}
P^{(2)}_1&=& -c_3[1] - d_3[1] - d_1 (c_2 [1] - d_2 [1]) =  \nonumber \\
&=& - c_4[0,1] - d_4[0,1] - d_1(c_3[0,1] - d_3[0,1]) = \dots  =  \nonumber 
\\ \nonumber
&=& - c_n[\underbrace{0,\ldots,1}_{n-2}] -d_n[\underbrace{0,\ldots,1}_{n-2}] 
 - d_1 ( c_{n-1} [\underbrace{0,\ldots, 1}_{n-2} ] - 
d_{n-1}[\underbrace{0,\ldots, 1}_{n-2} ]) = \\ 
&=& \bigg(\frac{397}{96}+\frac{17}{2}\zeta_3-15\zeta_5\bigg) {\rm  C_F^2}
\label{k_401-d_401} 
\end{eqnarray}
But to obtain analytic expressions for $P_1^{(3)}$ 
and $P_2^{(2)}$ using the same $\beta$-expansion method, we must find 
the $\beta$-expansion
representations for the fourth-order coefficients in the PT series for the 
functions $D(a_s)$ and $C^{Bjp}(a_s)$.

This can be done after additionally evaluating the gluino contributions to 
these important quantities
analytically in the fourth PT order and taking the three-loop gluon 
effects in the QCD RG $\beta$-function evaluated in
the ${\rm \overline{MS}}$- scheme in \cite{50} into account.
The relations obtained above allow deriving a new theoretical expression 
for the sum $d_4 + c_4$ of the fourth-
order coefficients of the PT series. For this, we fix the number 
$n_f$ of fermion flavors from the condition $\beta_0({\rm n_f} = n_0)
= 0$ which corresponds to the Banks-Zaks ansatz \cite{51} and leads to the value 
$T_F n_0 = (11/4)C_A$. In this case, we
obtain
\begin{eqnarray}
 c_ 4(n_0) + d_4(n_0)& =& c_4[0] + d_4[0] + \beta_2(n_0)(c_4[0,0,1] + 
d_4[0, 0,1]) + \nonumber\\ 
&&+ \beta_1(n_0)(c_4[0,1] + d_4[0,1]) \label{eq:check1}
\end{eqnarray}
The terms in the right-hand side of (\ref{eq:check1}) are known from (\ref{eq:k4-d4}) 
and (\ref{eq:k_n1-d_n1}) 
(i.e.,$-c_4[0,0,1] - d_4[0,0,1]$) and from (\ref{k_401-d_401})
(i.e., $-c_4[0,1] - d_4[0,1]$). 
Substituting the value $n_0$ fixed above in $\beta_1$ and $\beta_2$ 
and using (\ref{eq:check1}), we obtain
\begin{eqnarray} 
\label{BZ}
{\rm d_4}(n_0) + {\rm c_4}(n_0)&=& %\\ 
%&&
-\frac{333}{1024}{\rm  C_F^4}+{\rm C_A C_F^3} \left(-\frac{1661}{3072}+
\frac{1309}{128}\zeta_3-
\frac{165 }{16}\zeta_5\right)  \nonumber \\
&&+{\rm C_A^2 C_F^2}  \left(-\frac{3337}{1536}+\frac{7}{2} \zeta_3-
   \frac{105}{16} \zeta_5\right)+
  {\rm  C_A^3  C_F} \left(-\frac{28931}{12288}+\frac{1351}{512}
   \zeta_3\right).~~~
\end{eqnarray}    
Fixing the number ${\rm n_f} = n_0$ of quark flavors in the concrete analytic 
expression $d_4(n_0) + c_4(n_0)$ following from the
calculations in \cite{33}, we find agreement with the right-hand side of 
(\ref{BZ}).

In summary, using the new representation of the generalized Crewther relation 
derived here (see Eq. (\ref{KM1})) and
also the $\beta$-expansion method in \cite{43} and the Banks-Zaks ansatz 
\cite{51} 
allowed obtaining an additional argument for
the correctness of the results of complicated and lengthy computer analytic 
calculations performed by a group from
the Institute for Nuclear Research, the Institut fur Theoretische 
Teilchenphysik (Karlsruhe), and the Skobeltsyn
Institute of Nuclear Physics (Moscow State University) \cite{33}. 
Moreover, the absence of transcendental terms
proportional to $\zeta_7$ and $\zeta_3^2$ from the right-hand side of (\ref{BZ}) 
after the $\beta_0$   coefficient vanishes confirms the
observation made in \cite{33} that such contributions to the coefficients 
$d_4$ and $c_4$ determined in the ${\rm \overline {MS}}$ scheme are
proportional to the first coefficient $\beta_0$ of the QCD RG $\beta$-
function (see the results in \cite{33} and expression (\ref{eq:c_4})).

{\bf Acknowledgements.}
%%%%%%%%%%%%%%%%%%%%%%%%%%%%%%%%%%%%%%%%%%%%%%%%%%%%%%%%%%%%%%%%%%%%%%% 
This work was reported at the XVI International Seminar on High Energy Physics
``Quarks-2010" (6-12 June 2010, Kolomna) and at the International Workshop Hadron Structure and QCD: From
Low to High Energies ``HSQCD 2010" (5-9 July 2010, Gatchina). The authors are grateful to the organizers of
those seminars for the invitations, to K. G. Chetyrkin and D. I. Kazakov for the useful questions and 
discussions, to D. Broadhurst, A. G. Grozin, and O. V. Teryaev for the interest in these studies, and to 
S. Brodsky, R. Crewther, and P. Minkowski for the support and the comments on the literature after the 
electronic preprint appeared. 
One of the authors (A. L. K.) also acknowledges the interest and careful 
study of this work by A. V. Garkusha.
This work was supported by the Russian Foundation for Basic Research (
Grant No. 11-01-00182).
%%%%%%%%%%%%%%%%%%%%%%%%%%%%%%%%%%%%%%%%%%%%%%%%%%%%%%%%%%%%%%%%%%%%%%%

\end{document}